# Fluctuation Superconductivity in Mesoscopic Aluminum Rings

Nicholas C. Koshnick[1], Hendrik Bluhm[1], Martin E. Huber[2], Kathryn A. Moler[1]*
1 Department of Applied Physics, Stanford University,
Stanford, CA 94305, USA
2 Department of Physics, University of Colorado at Denver and Health Sciences Center,
Denver, CO 80204, USA
*To whom correspondence should be addressed; E-mail: kmoler@stanford.edu.

**Fluctuations are important near phase transitions, where they can be difficult to describe quantitatively. Superconductivity in mesoscopic rings is particularly intriguing because the critical temperature is an oscillatory function of magnetic field. There is an exact theory for thermal fluctuations in one-dimensional superconducting rings, which are therefore expected to be an excellent model system. We measure the susceptibility of many rings, one ring at a time, using a scanning SQUID that can isolate magnetic signals from seven orders of magnitude larger background applied flux. We find that the fluctuation theory describes the results and that a single parameter characterizes the ways in which the fluctuations are especially important at magnetic fields where the critical temperature is suppressed.**

Superconductivity requires both electron pairing and the coalescence of the pairs into a macroscopic quantum state with long-range phase coherence, usually described as a single wavefunction. In restricted geometries, thermal energy allows contributions from multiple wavefunctions to dramatically change the behavior of the system (*1-3*). Experimental knowledge of such fluctuations in one dimension (1D) is largely derived from transport measurements (*4*), which require electrical contacts and an externally applied current. We use a contact-less technique to study fluctuation effects in isolated, quasi-1D rings in the temperature range where the circumference is comparable to the temperature-dependent Ginzburg-Landau (G-L) coherence length, $\xi(T)$.

Using a scanning micro-Superconducting QUantum Interference Device (SQUID), we detect the current in many individual quasi-1D aluminum rings, paying particular attention to small currents near each ring's superconducting transition temperature, $T_c$. Such measurements have many advantages. In 1D, the current about a ring, $I$, is related to the free energy, $F$, via $I = -\partial F / \partial \Phi_a$, where $\Phi_a$ is the flux through the ring from an applied magnetic field. Measuring $I$ as a function of $\Phi_a$ thus tests a fundamental thermodynamic variable and our understanding of the ring's state. If there are superconducting pairs that are coherent about the ring's circumference $L$, the ring's current near zero applied field is proportional to the density of pairs. Deviations from this mean field solution provide information about amplitude and phase fluctuations in the ring.

The mean field G-L solution predicts that the current near zero field should decrease linearly to zero as the temperature, $T$, approaches $T_c$. For small rings, we find a measurable current above $T_c$, a clear signature of fluctuations. The quasi-1D geometry allows a full numeric solution of thermal fluctuations in a G-L framework that includes non-Gaussian effects (*5,6*). Previous results on a single ring at zero applied field (7) disagreed strongly with that theory. We studied fluctuations in 15 rings, and found that 13 rings agree quantitatively with a full numeric solution, which was numerically intractable for the other 2 rings (supporting online text).

The results in an applied field are particularly interesting. Little and Parks (*8*) showed experimentally that $T_c$ varies as a periodic function of $\Phi_a$, $T_c(\Phi_a)$. At half-integer multiples of the superconducting flux quantum, $\Phi_0$, the energetic cost of maintaining the flux-induced supercurrent can be larger than the condensation energy, destroying superconductivity. Previous results (*9-11*) indicate qualitatively that fluctuations are especially important in this regime. We find an enhanced response at $\Phi_a = \Phi_0/2$ that can be quantitatively explained by G-L thermal fluctuations and demonstrate that a single parameter can characterize the Gaussian and non-Gaussian regimes, and determines where the Little-Parks effect is entirely washed out by fluctuations.

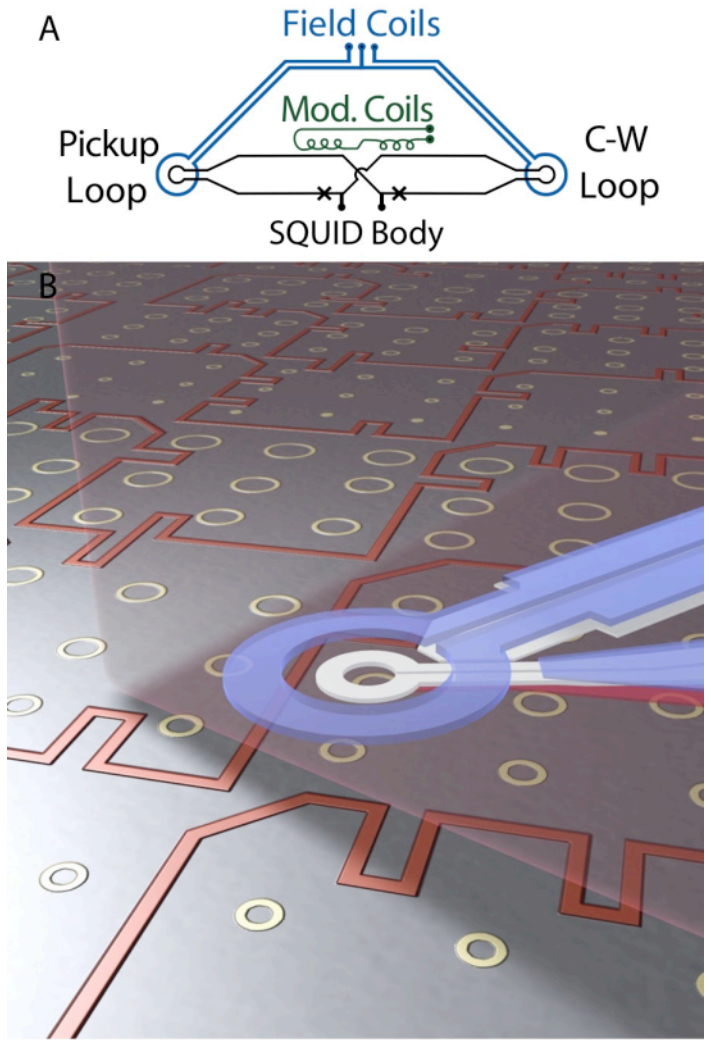

Fig 1. **A**: Diagram of the DC SQUID susceptometer. One field coil applies up to 50 Gauss of field to the sample, whose response couples a magnetic flux into the 4 μm pickup loop. A second counter-wound (C-W) loop cancels the SQUID's response to the applied field to within one part in $10^4$. Additional modulation coils maintain the optimum working point. **B**: The SQUID's pickup loop (white) and field coil (blue) are positioned over a single micron-scale aluminum ring. In-situ background measurements allow the magnetic flux induced by currents in the ring to be unambiguously distinguished from the applied field, which is up to seven orders of magnitude larger.

Unlike stationary sensors, a scanning sensor (Fig. 1) can measure many samples during a single cool-down. We report measurements on rings with radii $R$ = 0.35, 0.5, 1, and 2 μm, annulus widths $w$ from 65 to 180 nm, and thickness $d$ = 60 nm. The scanning SQUID also allows excellent background cancellation (*12*). After background subtraction, the signal (Fig 2) is proportional to current in the ring.

Many of the features in Fig 2 can be explained with the mean field response obtained by the minimization of a 1D G-L functional. By assuming a homogenous line width of negligible width, this process gives the ring current (*7*),

$$\begin{aligned} I_n &= -\frac{wd\Phi_0}{\lambda(T)^2 L\mu_0}\left(\Phi_a/\Phi_0 - n\right)P \\ P &= \max\left(1 - \frac{\xi(T)^2}{R^2}\left(\Phi_a/\Phi_0 - n\right)^2, 0\right) \end{aligned} \tag{1}$$

where $1/\lambda(T)^2$ represents the superfluid density and n is the phase winding number imposed by the single-valuedness of the macroscopic wave function. The different n states result in periodic $I$-$\Phi_a$ curves. The $\Phi_a$-linear term is the London response where $1/\lambda(T)^2 \propto T_c(0) - T$ below $T_c(0)$ in the temperature range of interest, and 0 above $T_c(0)$. P describes pair-breaking due to an Aharonov-Bohm phase around the ring, which leads to a downturn of the response at finite field when $\xi(T) >\sim R$. In small rings, this effect occurs well below $T_c(0)$ (Fig. 2A-C). The Little-Parks effect occurs in the temperature range where $\xi(T) \geq 2R$, bringing P to zero for a range of applied flux. The dashed green line in Figs 2C, D shows the best match to eq. 1 at 1.22K. The data's large remnant response in the region in which the mean field curve vanishes is a clear demonstration that fluctuation effects are important in this regime. In large rings (Fig 2D), fluctuations dominate the response before the effect of the pair-breaking term is apparent.

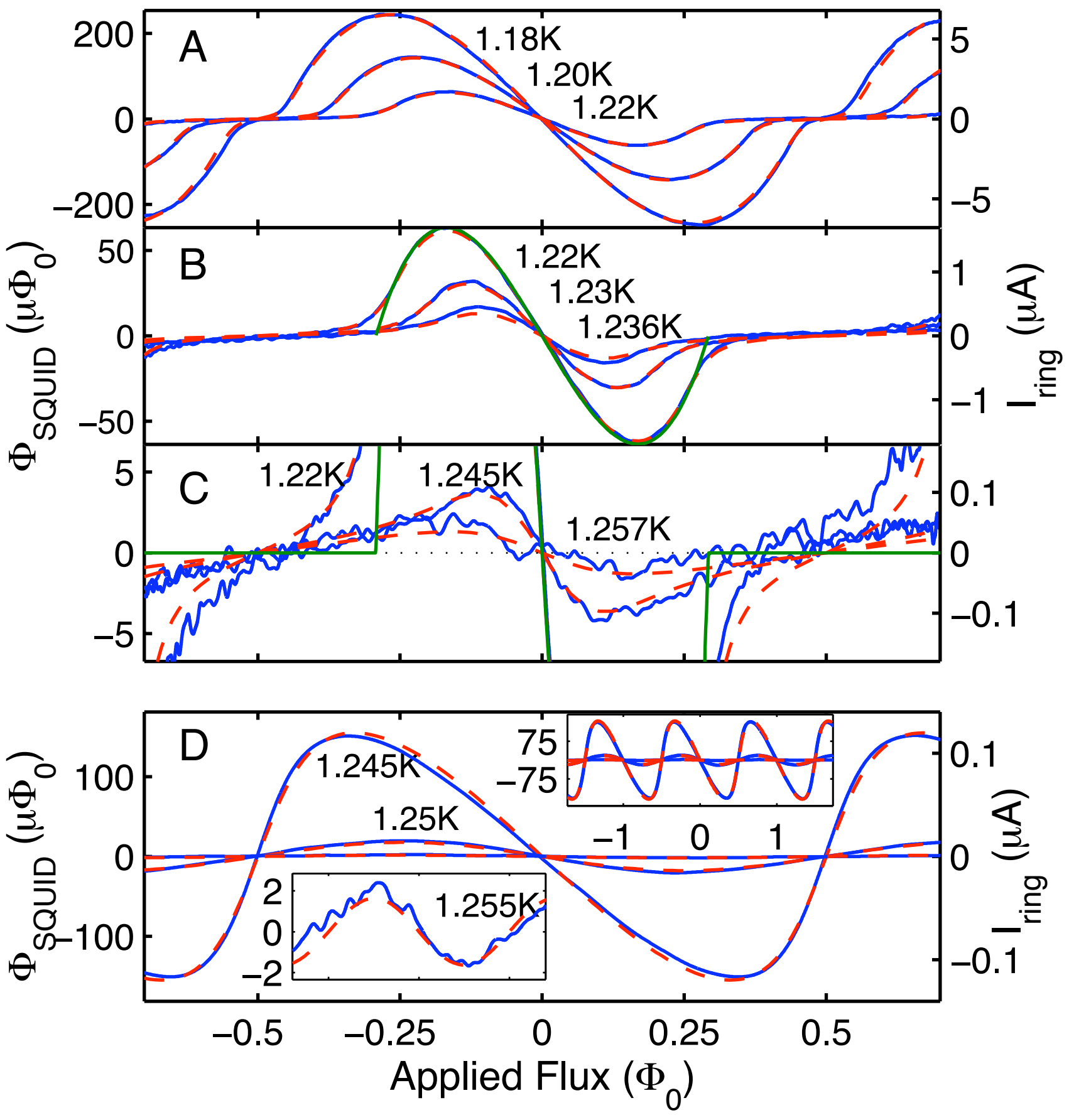

Fig 2. SQUID signal (left axis) and ring current (right axis) as a function of applied flux $\Phi_a$ for two rings, both with thickness d = 60 nm and width w = 110 nm. The fluctuation theory (red) was fit to the data (blue) through the temperature analysis shown in Fig 3. **A-C** Radius R = 0.35 μm, fitted $T_c(\Phi_a=0)$=1.247 K, and γ = 0.075. The green dashed line is the theoretical mean-field response for T = 1.22K and shows the characteristic Little-Parks lineshape, in which the ring is not superconducting near $\Phi_a = \Phi_0/2$. The excess persistent current in this region indicates the large fluctuations in the Little-Parks regime. **D** Radius R = 2 μm, fitted $T_c(\Phi_a=0)$=1.252 K, and γ=13. The periodic response (right inset) shows 1D treatment is appropriate and can be approximated by a thermal average over mean field G-L fluxoid states (Eq 1, supporting online text) until additional fluctuations contribute near $T_c$.

In the theoretical framework of (*5*), the current is given by $I \equiv -\partial F / \partial\Phi_a = k_B T \frac{\partial}{\partial\Phi_a} \ln Z_{sc}$, where the superconducting partition function, $Z_{sc}$, is the path integral over all possible wavefunctions. As shown in (*5*,*6*, supporting online text), $Z_{sc}$ can be written as $Z_{sc} = \sum_{l=-\infty}^{\infty} \exp\left(-i2l\Phi/\Phi\right) \sum_{n=0}^{\infty} \exp\left(-\gamma^{1/3} E_{n,l}\right)$ where $E_{n,l}$ are the eigenvalues of the 2D single-particle Hamiltonian, $H = -\frac{1}{2}\nabla^2 + \frac{4\pi}{\gamma^{2/3}} \frac{T - T_c(0)}{E_c / k_B} \vec{r}^{\,2} + \frac{1}{4}\vec{r}^{\,4}$. This formulation allows a numerical calculation of I that is exact up to truncation errors. We emphasize the parameter

$$\gamma \equiv \frac{42\zeta(3)}{\pi} \frac{k_B T_c}{M_{eff} E_c} \approx \frac{.87}{M_{eff}} \frac{L^2}{\ell_e \xi_0}, \qquad (2)$$

where $E_c = \pi^2 \hbar v_F \ell_e / 3L^2$ is the correlation or Thouless energy, ζ is the Riemann zeta function, $v_F$ is the Fermi velocity, and $\xi_0$ is the Pippard zero temperature coherence length. The annulus width w and thickness d enter into the effective number of channels, $M_{eff} = (\ell_e / L)(k_F^2 wd / 4\pi)$, where $k_F$ is the Fermi wavelength. This combination of parameters characterizes the size of the ring. Temperature only appears in the second term of the Hamiltonian, which can be rewritten using $8\pi k_B (T - T_c(0)) / E_c = L^2 / \xi(T)^2$ to illustrate the relation to the pair-breaking term of Eq. 1, and indicate the region where $T > T_c(\Phi_0/2)$ in Fig 3C. Thus, once the correct $E_c$ and $T_c(0)$ are known for a given ring, the current as a function of $\Phi_a$ and $T$ is entirely determined by γ.

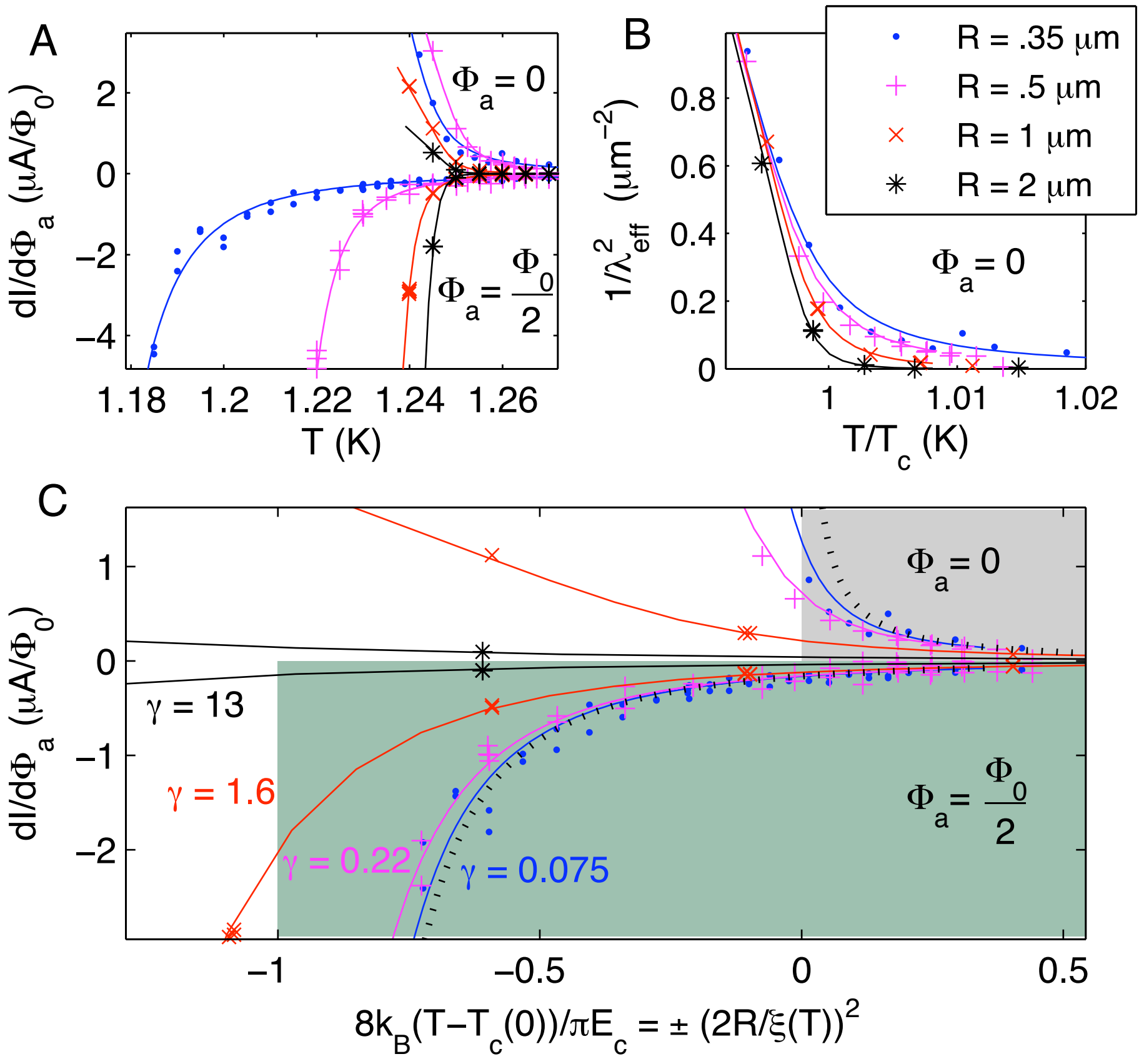

Fig 3. Susceptibility data (symbols) and fits (lines) at $\Phi_a = 0$ (positive values) and $\Phi_0/2$ (negative values) for 110 nm wide 60 nm thick rings with various radii, R. **A** Smaller rings have a larger temperature region where the Little-Parks criterion $\xi(T) > 2R$ is satisfied, and thus have a larger region with a reduced $\Phi_a = \Phi_0/2$ response. **B** $\Phi_a = 0$ susceptibility scaled with the cross section and radius to show the effective mean field superfluid density around $T_c$. Smaller rings have an enhanced fluctuation response. **C** When the temperature is scaled by the correlation energy $E_c$, the susceptibility is uniquely determined by the size parameter γ. The grey and green shaded regions indicate the temperature above $T_c(\Phi_a)$ when $\Phi = 0$ and $\Phi_0/2$, respectively. The fluctuation response above $T_c(\Phi_a)$ is enhanced for $\Phi_a = \Phi_0/2$. The dotted line shows a Gaussian prediction (supporting online text) that is valid at some γ-dependent temperature above $T_c(\Phi_a)$. When $\gamma >\sim 1$, the response at $\Phi_a = 0$ and $\Phi_0/2$ are comparable in the Little-Parks regime, which corresponds to a fluctuation-dominated sinusoidal I - $\Phi_a$ response.

The data points in Fig 3 were derived from I-$\Phi_a$ curves (e.g., Fig 2) by fitting low order polynomials near $\Phi_a = 0$ and $\Phi_a = \pm \Phi_0/2$. We have compared these susceptibilities as a function of temperature to the theory by using the measured geometry factors, the $k_F$ and $v_F$ of bulk aluminum, and three parameters, chosen by hand: $\ell_e$, $T_c(0)$, and $M_{S\text{-}R}$, the mutual inductance between the SQUID and the ring. $\ell_e$ is identified by the shape of the curve as a function of temperature, $M_{S\text{-}R}$ is determined by the magnitude of the response, and $T_c(0)$ is chosen to allow the theory to match the linear temperature dependence of the $\Phi_a = 0$ susceptibility below $T_c(0)$. The $T_c(0)$ of the rings varied from 1.237 to 1.268 K with up to 7 mK difference for nominally identical line-width rings. The fitted $M_{S\text{-}R}$ lie within 15% of the inductance calculated with a model based on a 0.75 μm ring -- sensor-loop separation. The fitted $\ell_e$ varies between 14 and 25 nm with an increasing dependence on line-width. We attribute this dependence to oxygen absorbed during the fabrication process and, to a lesser extent, the observed 20% variations in *w* (*12*). The 4 nominally identical line-width rings shown in Fig 3 have $\ell_e = 19.5 \pm 0.5$ nm. The agreement with the 1D models we have discussed demonstrates that the finite line-width effects are not essential to our physical result and that small variations in w do not qualitatively change the response above $T_c(\Phi_a)$.

Near $T_c(\Phi_a)$, $\gamma$ characterizes the non-Gaussian fluctuations that interpolate between the mean field behavior far below $T_c$ and the Gaussian fluctuations that dominate at high temperatures. Non-Gaussian fluctuations are important when quadratic expansions of the free energy cannot describe the physical result. This is particularly apparent at $T_c(\Phi_a)$, where any Gaussian approximation would predict a divergent susceptibility (Fig 3A). By using Eq 1 to define an effective superfluid density from the zero field response, $1/\lambda_{eff}^2 = (\mu_0 L / wd)\partial I / \partial \Phi_a|_{\Phi_a=0}$, one can see (Fig 3B) that fluctuations make the susceptibility deviate from the mean field response below $T_c(0)$, gradually smoothing the transition. Our parameterization of the theory shows that $\gamma$ is the only sample dependent parameter at $T = T_c(0)$. The temperature range where non-Gaussian fluctuations are important is typically parameterized through the Ginzburg parameter as $|T-T_c(\Phi_a)|/T_c(\Phi_a) < Gi$, where $|T - T_c(\Phi_a)|/E_c < \frac{T_c(\Phi_a)}{E_c} Gi \propto \sqrt{\gamma}$ (*13*). Inside this range, $\gamma$ determines the magnitude of the response. Far above this range, Gaussian fluctuations dominate and the susceptibility is a function of $|T-T_c(0)|/E_c \propto L^2/\xi^2$ alone.

The theory's dependence on $\gamma$ allows us to state the criterion for the visibility of the Little-Parks effect in the context of fluctuations. The region that is shaded in green in Fig 3C is above $T_c(\Phi_0/2)$ because $\xi(T) > 2R$. The susceptibility would be zero in this regime if fluctuation effects were not considered. When $\gamma \ll 1$, the distinct Little-Parks shape is visible, in that the susceptibility is smaller at $\Phi_a = \Phi_0/2$ than at $\Phi_a = 0$. However, when $\gamma \gg 1$, the Little-Parks shape is entirely washed out by fluctuations (Fig 4). For sufficiently large $\gamma$, the susceptibilities at $\Phi = 0$ and $\Phi_0/2$ are equal and opposite even below $T_c(\Phi_a = \Phi_0/2)$ so the response appears sinusoidal. This dependence on $\gamma$, rather than $L$ and $\xi$(T) alone, is the reason why the Little-Parks lineshape does not occur in the ring shown in Fig 2A, 4C.

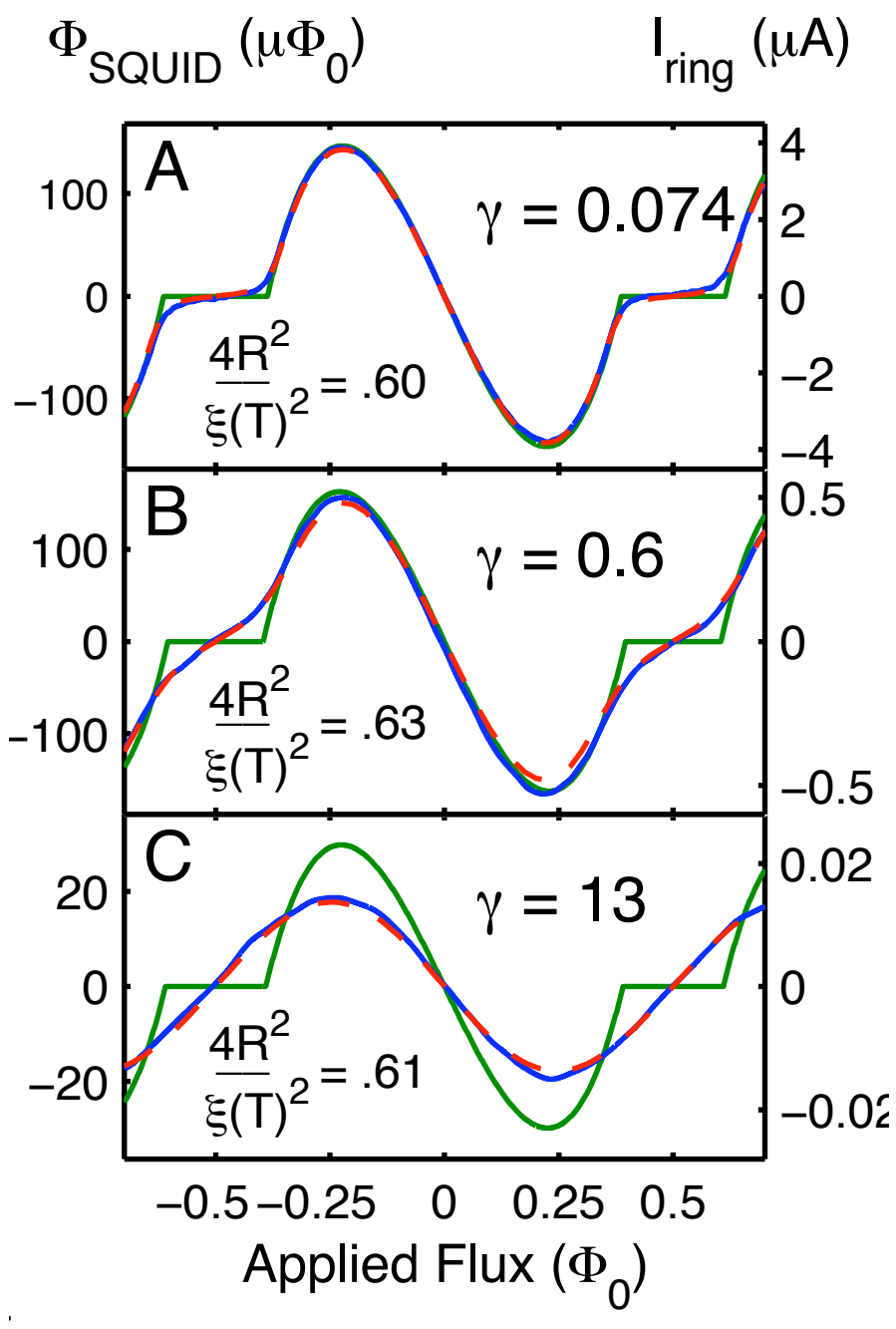

Fig 4. Mean field theory (green), fluctuation theory (red) and data (blue) for three rings with different $\gamma$ parameters. The mean field response is derived from the fluctuation theory parameters for each ring at the given temperature. **A** T = 1.20K. In small $\gamma$ rings, the Little-Parks line shape is clearly observable. **B** T = 1.25K. When $\gamma \approx 1$, the reduction of the response due to the Little-Parks effect is significantly suppressed. **C** T = 1.25K. In large $\gamma$ rings, the Little-Parks effect is completely washed out by fluctuations, which affect the response at all flux values.

Several factors contribute to the large fluctuation response near $\Phi_a = \Phi_0/2$ above $T_c(\Phi_a)$. First, the Gaussian fluctuations between $T_c(\Phi_a = 0)$ and $T_c(\Phi_a)$ have a large magnitude which is due to the interplay between adjacent phase winding states. In small $\gamma$ rings, the non-Gaussian fluctuation region in Fig 3C is small. Thus, there is a large region where the magnitude of the persistent current near $\Phi_a = \Phi_0/2$ is strictly a function of $k_B(T\text{-}T_c(0))$ /$E_c$. In large $\gamma$ rings, non-Gaussian fluctuations play an increased role in the phase diagram and multiple phase winding modes need to be considered (*13*), indicating the importance of phase fluctuations. In all rings, non-homogeneous wavefunctions may have a non-negligible contribution to the final currents due to their vanishing energy cost near $T_c(\Phi_a)$. Small variations in width (supporting online text) make non-homogeneous wavefunctions more important (*14*), and would be important to include in an extended theory.

Fluctuation effects play a important role in 1D superconducting structures. Our analysis explicitly demonstrates how Gaussian and non-Gaussian fluctuations affect the persistent current in rings with various diameters and cross-sections, as a function of applied magnetic flux. A single parameter, $\gamma$, characterizes the fluctuations for a given ratio of the temperature-dependent coherence length to the circumference. When $\gamma$ is large, the signature of a Little-Parks flux dependent $T_c(\Phi_a)$ is entirely washed out by fluctuations. When $\gamma$ is small, the susceptibility in the non-Gaussian region near $T_c(\Phi_a)$ is enhanced and Gaussian fluctuations are clearly visible between $T_c(\Phi_a)$ and $T_c(0)$ for $\Phi_a \approx \Phi_0/2$. This new framework for understanding Little-Parks fluctuations is supported by our data on fluctuation-induced currents in rings.

15. We acknowledge support from the Packard Foundation and NSF Grants No. DMR-0507931, DMR-0216470, ECS-0210877, ECS-9731293 and PHY-0425897. We would like to thank J. Price, Y. Imry, M.R. Beasley, Y. Oreg, and especially G. Schwiete for helpful discussions.

Supporting Online Material
www.sciencemag.org
Materials and Methods
Supporting Online Text
Tables S1, S2